\documentclass[aps,prl,twocolumn,notitlepage,showpacs,superscriptaddress,am]{revtex4-1}%
\usepackage{graphicx}
\usepackage{amsmath}
\usepackage{amssymb}
\usepackage{color}
\usepackage{amsfonts}%
\providecommand{\U}[1]{\protect\rule{.1in}{.1in}}
\def\Herb{$\rm ZnCu_3(OH)_6Cl_2$}
\def\Y{$\rm Y_3Cu_9(OH)_{19}Cl_8$}

\begin{document}
\title{Nature of optical excitations in the frustrated kagome compound Herbertsmithite}
\author{A. Pustogow}
\affiliation{1.~Physikalisches Institut, Universit\"{a}t Stuttgart, 70569 Stuttgart, Germany}
\author{Y. Li}
\affiliation{Institut f\"{u}r Theoretische Physik, Goethe-Universit\"{a}t Frankfurt, 60438 Frankfurt am Main, Germany}
\author{I. Voloshenko}
\affiliation{1.~Physikalisches Institut, Universit\"{a}t Stuttgart, 70569 Stuttgart, Germany}
\author{P. Puphal}
\affiliation{Physikalisches Institut, Goethe-Universit\"{a}t Frankfurt, 60438 Frankfurt am Main, Germany}
\author{C. Krellner}
\affiliation{Physikalisches Institut, Goethe-Universit\"{a}t Frankfurt, 60438 Frankfurt am Main, Germany}
\author{I.I. Mazin}
\affiliation{Code 6393, Naval Research Laboratory, Washington, DC 20375, USA}
\author{M. Dressel}
\affiliation{1.~Physikalisches Institut, Universit\"{a}t Stuttgart, 70569 Stuttgart, Germany}
\author{R. Valent\'{\i}}
\affiliation{Institut f\"{u}r Theoretische Physik, Goethe-Universit\"{a}t Frankfurt, 60438 Frankfurt am Main, Germany}
\date{\today}

\begin{abstract}
Optical conductivity measurements are combined with density functional
theory calculations in order to understand the electrodynamic response of the frustrated Mott
insulators Herbertsmithite $\mathrm{ZnCu_{3}(OH)_{6}Cl_{2}}$\ and the
closely-related kagome-lattice compound $\mathrm{Y_{3}Cu_{9}(OH)_{19}Cl_{8}}$.
We identify these materials as charge-transfer
rather than Mott-Hubbard insulators, similar to the high-$T_c$ cuprate parent
compounds. The band edge is at 3.3 and 3.6 eV, respectively, establishing the
insulating nature of these compounds. Inside the gap, we observe
dipole-forbidden local electronic transitions between the Cu $3d$ orbitals in the
range 1--2 eV.  With the help of \textit{ab initio} calculations we demonstrate that
the electrodynamic response in these systems is directly related
to the role of on-site Coulomb repulsion: while charge-transfer processes
have their origin on transitions between the ligand band and the Cu $3d$
upper Hubbard band,  \textit{local}
$d$-$d$ excitations remain rather unaffected by correlations.

\end{abstract}

\maketitle

Since the discovery of high-$T_{c}$ cuprates, the physics of strongly
correlated materials has been at the forefront of research in condensed matter
physics. The relationship between correlations, unconventional
superconductivity, quantum-spin-liquid behavior and other exotic states of
matter has been intensely debated. While at the level of model theories one
can introduce criteria to quantify the degree of correlation, such as the
ratio $U/W$, where $U$ is the on-site Coulomb repulsion and $W$ is the
single-particle bandwidth, the quantification in many materials is not that straightforward. The effect of correlations usually shows up as mass
renormalization, band narrowing, or as opening or enhancement of the band gap~\cite{Basov2011,Imada1998}.

Cu$^{+2}$ ions are, arguably, the most strongly correlated among
$d$ transition metals~\cite{Zaanen1985,Zaanen1990,Bocquet1992,Olalde-Velasco2011}. They are key ingredients of
high-$T_{c}$ cuprates and  have been widely investigated in the last
decades. A recent revival of interest in  Cu-based materials was triggered by the discovery of
geometrically frustrated cuprates that seem to exhibit spin-liquid
properties~\cite{Helton2007,Lee2007,Mendels2007,DeVries2008,Lee2008,Balents2010,Mendels2010,Han2012,Norman2016,Savary2017} and may even harbor
unconventional superconductivity with higher angular momenta than existing
superconductors~\cite{Mazin2014} or further topological phases~\cite{Guterding2016} although synthesis seems to be difficult~\cite{Kelly2016}.

In this work we investigate the origin of the optical excitations in the
spin-liquid candidate Herbertsmithite and concentrate on the following
conceptual issue: which measurable properties in correlated systems are
strongly affected by correlation effects and which are not? We will show,
experimentally and theoretically, that in a single experimental probe, namely
optical conductivity, one can simultaneously observe properties dramatically
influenced by Coulomb (Mott-Hubbard) correlation effects, and those that are
hardly affected at all.

\begin{figure}[b]
\centering
\includegraphics[width=1\columnwidth]{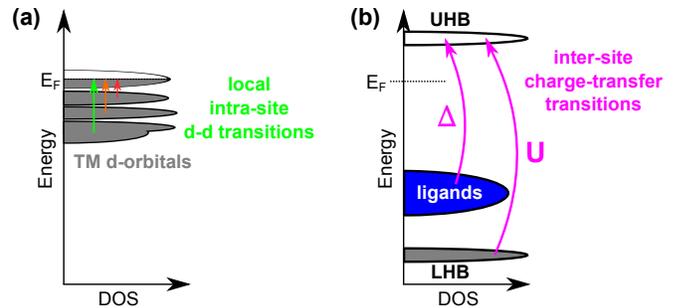}
\caption{(Color online) Two distinct types of inter-band excitations are
identified in strongly correlated transition-metal (TM) compounds. 
(a) \textit{Local} excitations between the $d$-orbitals on the same TM ion do 
not change the site charge and, hence, remain rather unaffected by electronic
correlations. (b) Charge transfer processes between different sites change 
the local electron density and are, therefore, strongly affected by the on-site
Coulomb interaction. While the charge-transfer gap $\Delta$ has to be overcome 
for excitation of an electron from the ligands to the upper Hubbard band (UHB) 
of the TM ion, the Coulomb repulsion $U$ has to be paid for inter-site TM ion - TM ion charge transfer, 
corresponding to transitions between the Hubbard bands.} 
\label{CT-processes}%
\end{figure}

One can distinguish two types of optical absorption processes, depicted in Fig.~\ref{CT-processes}. On the one hand, we have the
\textquotedblleft charge-transfer process\textquotedblright, that in its
simple form can be described as creating a hole and an electron residing at
different lattice sites; the corresponding changes of charge distribution modify the Coulomb energy. 
The second process, usually referred to as local $d$-$d$ transitions, creates an electron-hole
pair residing on the same transition-metal site which, nominally, does not enhance the Hubbard repulsion. The main conceptual problem with this
picture is that the notion, rigorously speaking, applies only to an isolated
transition-metal ion --- in which case any local transition between the $d$-orbitals is
forbidden by symmetry~\cite{Lever1968,Burns1993,Solomon1999}. One $can,$ however, meaningfully apply the same
nomenclature in an extended solid, but then it should refer to the
one-electron propagator in real space that originates and ends on the same
site, or on different sites. This way, any residual itinerancy can be
formally incorporated into the concept. Note, while photoemission probes
the self-energy of the single-particle Green function and is
described by diagrams changing the number of electrons, optical absorption
is a polarization bubble (self-energy of the two-particle Green function)
and, thus, always \textit{conserves} the total number of particles in the system
(but not necessarily on each site).

In this language, the problem of the dipole matrix elements solves itself
automatically. In an extended solid, a Green function originating at a given
transition-metal site is not necessarily  localized on
the same site. As long as the corresponding atomic state acquires dispersion,
it always includes some admixture of \textquotedblleft orbital
tails\textquotedblright\ coming from other sites, primarily ligands,
but not only. In density functional and similar calculations it is seen as admixture
of other characters into formally $d$-bands. Now it is important to recognize
that around each site, say, within the atomic sphere, electronic wave
functions, no matter whether \textquotedblleft native\textquotedblright\ or
penetrating from outside, have to solve the radial Schr\"{o}dinger equation
for the \textquotedblleft host\textquotedblright\ site. In other words, the
tails of surrounding orbitals are re-expanded around the transition-metal site as a
linear combination of the transition-metal $s$-, $p$-, $d$-, $f$- orbitals. Therefore, in
optical calculations the so-called $d$-$d$ excitations are treated as local, but
the corresponding local states have admixture of other characters; the dipole
transitions proceed between the main $d$ portion of the
angular-momentum-decomposed states and the minor admixture of $p$- and $f$- characters.

This is, in a nutshell, the physics we want to address.
 We have chosen
herbertsmithite-type compounds, $\mathrm{ZnCu_{3}(OH)_{6}Cl_{2}}$\ and
$\mathrm{Y_{3}Cu_{9}(OH)_{19}Cl_{8}}$\, as our testing ground, for the
following reasons: (i) they have, as we will show, sizeable Mott-Hubbard
correlation gaps of the order of 3 eV, and thus leave a window of transparency
at lower energies where we expect the $d$-$d$ excitations to manifest
themselves; (ii) while intensively studied by magnetic and thermodynamic probes\cite{Mendels2007,Helton2007,Lee2007,Olariu2008,DeVries2008,Zorko2008,DeVries2009,Helton2010,Han2012,Han2012a,Zorko2017,Puphal2017},
optical investigations focused only on the low-energy excitations and phonons~\cite{DeVries2012,Pilon2013,Sushkov2017}
and the details of the band structure of these materials have not
been well characterized; even such basic parameters, extremely important for
theoretical models, such as the Coulomb repulsion $U$ and the hierarchy of the
crystal field levels, have not been determined reliably in experiment so far.
Hence, our goal is twofold: to fill a lacuna in the characterization of these
intensely studied compounds and to gain insight into the physics of the
local $d$-$d$ optical excitations.

\begin{figure}[ptb]
\centering
\includegraphics[width=1\columnwidth]{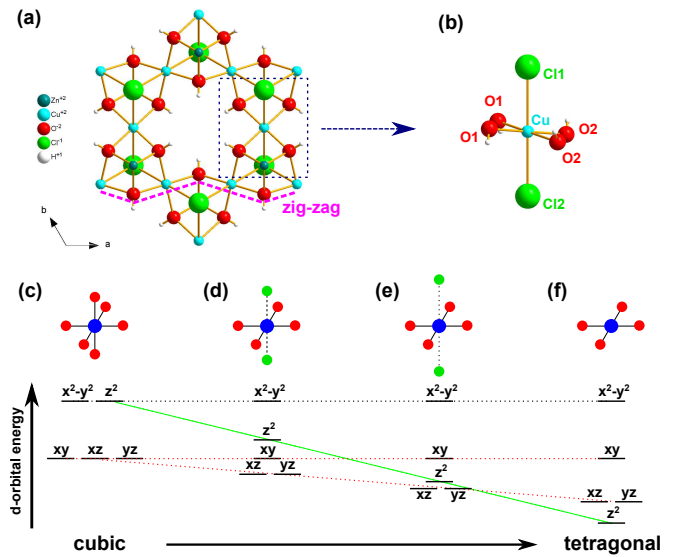}
\caption{(a) The crystal structure of $\mathrm{ZnCu_{3}%
(OH)_{6}Cl_{2}}$ exhibits the characteristic kagome arrangement of Cu atoms~\cite{Shores2005}.
(b) The $\mathrm{CuO_{4}Cl_{2}}$
octahedra are strongly distorted as the Cl atoms are more distant from the
center Cu site than the O atoms. In addition, the CuO$_{4}$ tetragon is tilted
with respect to the equatorial plane perpendicular to the Cl-Cu-Cl axis, effectively flattening the zig-zag shape of the CuO chains with respect to the kagome layer. (c)-(f)
Increasing the vertical distortion lowers the $z^{2}$ orbital relative to
the $x^{2}-y^{2}$ state~\cite{MorettiSala2011}.}
\label{structure}
\end{figure}

Single crystals of $\mathrm{ZnCu_{3}(OH)_{6}Cl_{2}}$ and $\mathrm{Y_{3}
Cu_{9}(OH)_{19}Cl_{8}}$ were grown by hydrothermal methods. We reproduced the
hydrothermal synthesis of several batches of Herbertsmithite single crystals following Ref.~\onlinecite{Han2011}.
$\mathrm{Y_{3}Cu_{9}(OH)_{19}Cl_{8}}$\ single crystals were synthesized
hydrothermally following Ref.~\onlinecite{Puphal2017} using a duran glass
ampoule in an autoclave heated to ${270}^\circ$~C followed by slow cooling to ${260}^\circ$~C. The optical reflectivity was measured on thick  ($d>0.4$~mm) samples, and for
transmission measurements very thin ($10$ ${\rm \mu m}\lesssim d\lesssim70$ ${\rm \mu
m}$), plate-like crystals were selected. The experiments were performed with a
Woollam ellipsometer covering the frequency range 0.6 eV $<\hbar\omega<$ 6 eV. The
transmission was probed at normal incidence only, while for ellipsometry
reflection geometry with various angles of incidence was employed. In the
visible range and at lower frequencies we performed additional measurements at
normal incidence in a commercial Bruker Fourier-transform infrared spectrometer (0.005 eV $<\hbar\omega<$
3 eV), confirming the transmission data in the overlapping range. We also
calculated the optical conductivity from the broadband reflectivity using the
Kramers-Kronig relation, verifying our ellipsometric results~\cite{SM}.

\begin{figure*}[ptb]
\centering
\includegraphics[width=1.5\columnwidth]{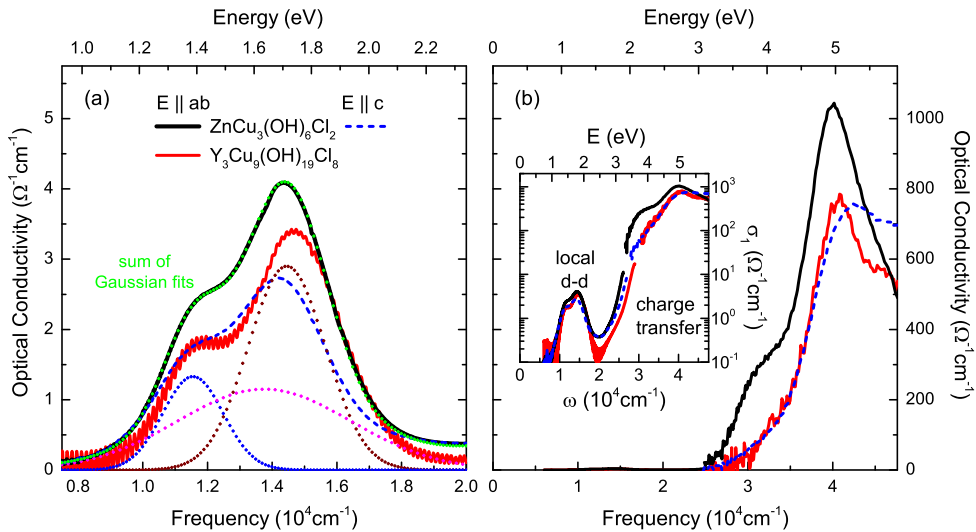}
\caption{The optical conductivity of $\mathrm{ZnCu_{3}(OH)_{6}Cl_{2}}$ (solid black) and
$\mathrm{Y_{3}Cu_{9}(OH)_{19}Cl_{8}}$ (solid red) in the visible and ultraviolet range
shows general similarity proving that the electrodynamics is mainly determined
by the Cu kagome network. (a) We assign the weak low-frequency
multi-peak to dipole-forbidden local transitions between the Cu $3d$ orbitals.
One broad and two narrow Gaussians are necessary to satisfactorily fit the
data (small symbols). For $\mathrm{ZnCu_{3}(OH)_{6}Cl_{2}}$\ crystal geometry also allowed to
measure the out-of-plane conductivity (dashed blue), which has similar frequency dependence,
but smaller intensity. (b) The strong high-frequency feature is identified as
the charge transfer between ligands and the unoccupied Cu $3d$ states (upper Hubbard band). The inset illustrates the different intensity of local and charge-transfer processes on a logarithmic scale.}%
\label{sigma_VIS-UV}%
\end{figure*}

For the density functional theory (DFT) calculations we used the
experimentally determined crystal structure of $\mathrm{ZnCu_{3}(OH)_{6}Cl_{2}}$~\cite{Shores2005}, and employed
the linearized augmented plane-wave (LAPW) package WIEN2k~\cite{Blaha2001}
with the generalized gradient approximation (GGA)~\cite{Perdew1996}. The
basis-controlling parameter was set to $RK_{\rm max}=3$ due to the presence of H
in the structure. This value is equivalent to $RK_{\rm max}\sim7-9$ for oxides. We
used a mesh of 1000~$\mathbf{k}$ points in the first Brillouin
zone (FBZ) of the primitive unit cell. The density of states (DOS) and hopping parameters were computed with $20\times
20\times20~\mathbf{k}$ points in the full Brillouin zone for the GGA calculation. In order to address
correlation effects in the charge-transfer processes, we used the GGA + $U$ method
in the spherically averaged approximation~\cite{Anisimov1993} and assumed
various antiferromagnetic configurations (see Ref.~\onlinecite{SM}). For DOS and optical conductivity GGA + $U$ calculations, we have chosen 2340 $\mathbf{k}$ points in the primitive unit cell.

Fig.~\ref{structure} shows the details of the Herbertsmithite structure. Each
Cu$^{+2}$ is surrounded by four O$^{-2}$ and two more distant Cl$^{-}$
anions, forming a distorted octahedron [Fig.~\ref{structure}(b)]. The interatomic distances
are listed in Ref.~\onlinecite{SM}. Chemical disparity
between Cl and O implies a considerable tetragonal component in the crystal field. The Cu-Cl bond is significantly longer than the Cu-O one, even
accounting for the larger radius Cl$^{-}$. Hence, in the simple
ligand-field picture the $e_{g}$ levels will split as $x^{2}-y^{2}$ (high) and
$z^{2}$ (low), and the  $t_{2g}$ states as $xy$ (high) and $\{xz,yz\}$ (low), as illustrated in
Fig.~\ref{structure} (c)-(f). In a real crystal, however, the arrangement of $d$-orbitals is hard to predict without actual calculations.
Another effect of stretching the CuO$_{4}$Cl$_{2}$
octahedra is the tilt of the O$_{4}$ tetragon with respect to the
equatorial plane perpendicular to the Cl-Cu-Cl axis. This tilting effectively
straightens the Cu-O zig-zag chains oriented along $a$, $b$ and
$a+b$ and affects the bandwidth and band gap.

Fig.~\ref{sigma_VIS-UV} displays the optical conductivity of $\mathrm{ZnCu_{3}%
(OH)_{6}Cl_{2}}$ and $\mathrm{Y_{3}Cu_{9}(OH)_{19}Cl_{8}}$ measured in the
visible and ultraviolet ranges. Two distinct features are
identified in the spectra of both compounds as it is common in Cu$^{+2}$ systems~\cite{Olalde-Velasco2011,MorettiSala2011,Riley1990,Perkins1993,Bassi1996,Perkins1998,deGraaf2000,Hwu2002,Pisarev2010,Pisarev2011}. 
While local d-d transitions appear around 1--2~eV, intense charge-transfer excitations set in above 3 eV, where the original hole of
the Cu ${x^{2}-y^{2}}$ state is filled by an electron from the ligands, creating a hole on the ligands. To reproduce the latter process in standard $ab$
$initio$ calculations one has to include the effect of the Hubbard repulsion,
which is done in the simplest mean-field way in the framework of the GGA + $U$ formalism.

The optical conductivity shows an interesting structure: the
absorption edge is around 3.3 and 3.6 eV for $\mathrm{ZnCu_{3}(OH)_{6}Cl_{2}}%
$\ and $\mathrm{Y_{3}Cu_{9}(OH)_{19}Cl_{8}}$, respectively. This difference is likely due
to different CuO$_{4}$ tiltings -- \Y\ is closer to a rectangular arrangement 
than \Herb ~\cite{SM} -- as discussed above. Contrasting the in- and out-of-plane polarizations of $\mathrm{ZnCu_{3}(OH)_{6}Cl_{2}}$, we find significant anisotropy.

\begin{figure}[ptb]
\centering
\includegraphics[width=0.8\columnwidth]{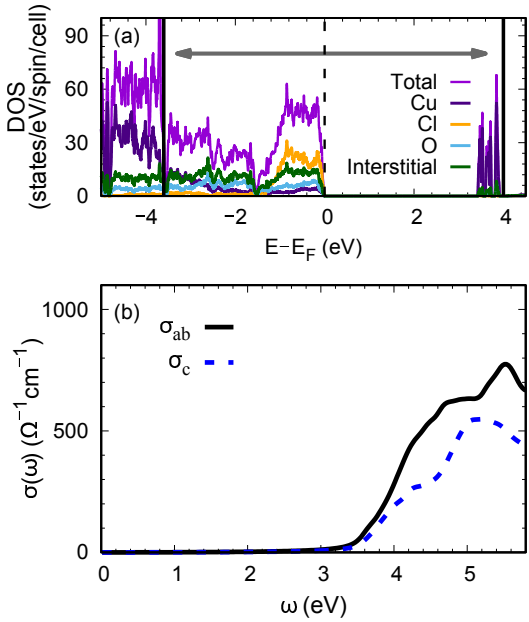}
\caption{(a) Density of states (DOS) and (b) optical conductivity for in- ($\sigma_{ab}$) and out-of-plane ($\sigma_{c}$) directions of $\mathrm{ZnCu_{3}(OH)_{6}Cl_{2}}$ calculated by GGA + $U$. It reproduces the charge-transfer gap observed in experiment for $U_{\mathrm{eff}}=U-J_{H}=8$~eV. The arrow in (a) indicates the energy range we considered to calculate $\sigma(\omega)$ in (b). The gap of 3.4 eV and main peaks around 5 eV
coincide with the data in Fig.~\ref{sigma_VIS-UV} (b).}%
\label{theory_CT-gap}%
\end{figure}

In Fig.~\ref{theory_CT-gap} we present the DOS and optical
conductivity obtained from GGA + $U$ calculations with $U_{\mathrm{eff}}=U-J_{H}\approx 8$ eV which is a typical value for Cu$^{+2}$. $J_H$ defines the Hund's rule coupling, usually 0.9--1~eV for $3d$ electrons. These results reproduce well the optical transitions observed above 3 eV in
$\mathrm{ZnCu_{3}(OH)_{6}Cl_{2}}$ [Fig.~\ref{sigma_VIS-UV} (b)]. The dominant ligand character of the occupied states highest in energy in Fig.~\ref{theory_CT-gap}~(a) clearly establishes Herbertsmithite as a charge-transfer insulator with the spectral gap between the ligand band and the unoccupied
Cu $3d$ band. The large Cu contributions below $E-E_F \approx -4$~eV correspond to the lower Hubbard band. In
particular, we can identify the gap of 3.4 eV, a shoulder at 4 eV (4--4.6 eV
in the calculations) and the strong ascend at 5 eV (5--5.5 in the
calculations). Even the experimental anisotropy is well reproduced. In our
calculation, the antiferromagnetic configuration introduces in-plane anisotropy between
$\sigma_{yy}$ ($y$ is along $b$) and $\sigma_{xx}$ ($x$ is perpendicular to
$b$), we averaged these to obtain $\sigma_{ab}=(\sigma_{xx}+\sigma_{yy})/2$~\cite{SM}. The origin of the transitions can be traced back to the DOS as shown in Fig.~\ref{theory_CT-gap} (a). Note that in the LAPW calculations there are
considerable interstitial contributions to the DOS. The small
O-H distance requires the use of small RMT (1.01 \AA ) in the calculation, implying that oxygen states reach into the interstitials. We can identify the optical processes as
charge transfer transitions from ligands $p$ states to the Cu $3d$
upper Hubbard band. The arrow between -3.6 eV and 4 eV in Fig.~\ref{theory_CT-gap} (a) indicates the
energy region considered for the optical conductivity calculations.

\begin{figure}[ptb]
\centering
\includegraphics[width=0.8\columnwidth]{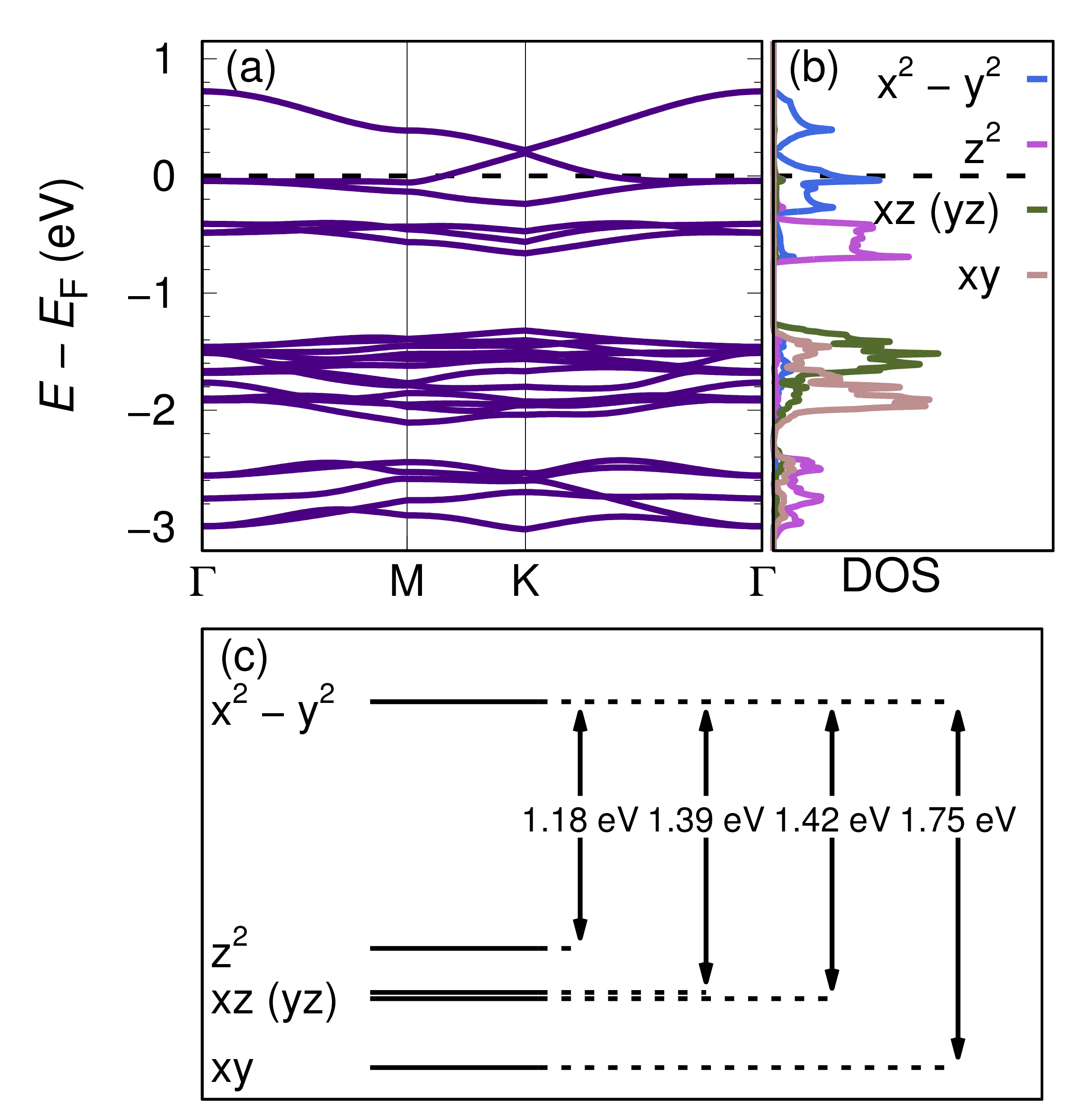}
\caption{ Calculated GGA
(a) band structure and (b) Cu $3d$ orbital resolved density of states (DOS) for
$\mathrm{ZnCu_{3}(OH)_{6}Cl_{2}}$ at $U=0$. (c) Energy level diagram for
on-site Cu $3d$ orbitals. The energy differences of the three levels 1.18 eV, 1.39
eV and 1.42 eV correspond to the experimental peaks in
Fig.~\ref{sigma_VIS-UV} (a).}%
\label{theory_d-d}%
\end{figure}

Now we turn our attention to the measured local \textit{dipole-forbidden d-d} low
energy excitations, displayed in Fig. ~\ref{sigma_VIS-UV} (\textit{a}), which
appear $inside$ the charge transfer gap (inset panel \textit{b}). They are visible 
in the experiment as a double peak around 1--2 eV causing the blue-green color of 
the $\mathrm{ZnCu_{3}(OH)_{6}Cl_{2}}$ and $\mathrm{Y_{3}Cu_{9}(OH)_{19}Cl_{8}}$ crystals.
The intensity is small since, as discussed above, these transitions are only
possible because the ligand orbitals penetrate the vicinity of the Cu site and
get re-expanded into the $l=1$ and $l=3$ harmonics. This is corroborated by
our calculations which reveal that the Cu $d$ states are mixed with O and Cl $p$
states (see Ref.~\onlinecite{SM}). The experimental data can be
satisfactorily fitted by three Gaussians, as indicated for $\mathrm{ZnCu_{3}(OH)_{6}Cl_{2}}$ by the blue, magenta and
wine curves in Fig.~\ref{sigma_VIS-UV} (a). In order to understand the origin
of these peaks, we need information of the crystal field splitting in the
extended crystal. This can be accurately obtained by GGA calculations as shown
in Fig.~\ref{theory_d-d} (a) and (b).
We obtain the following band order:
${x^{2}-y^{2}}$, ${z^{2}}$, ${xz}~({yz})$, ${xy}$. We find that while
the $xy$ and the $\{xz,yz\}$ states are very close in energy, their order is
reversed, the ${xy}$ being lower; this indicates that the simple
ligand-field model sketched in Fig.~\ref{structure} (c)-(f) is an oversimplification. We then calculate the on-site and
hopping parameters using the projector method described in
Refs.~\cite{Aichorn2009,Foyevtsova2013}. For that, due to the strong
hybridization of Cu with Cl, we consider 23 bands including Cu $d$, Cl $p$ and
Zn $e_{g}$ orbitals~\cite{SM,Kira}. The diagonalized on-site
hopping matrix elements after shifting the energy of the unoccupied ${x^{2}-y^{2}}$ state to
0 are $-1.18$ eV, $-1.39$ eV, $-1.42$ eV and $-1.75$ eV, with the dominant
characters as described above. We find overall a very good agreement of the level
positions with the multipeak structure of the observed local $d$-$d$ transitions.
We note that in the context of other transition-metal compounds, the origin of these transitions has been intensively debated, and other effects
than the above discussed mixing of states such as phonons, magnons and spin-orbit coupling have
been also suggested to contribute to making 'allowed' the dipole-forbidden $d$-$d$ transition~\cite{Lever1968,Burns1993,Solomon1999,Haverkort2007,Pisarev2011}. In our case
we can explain the main features in terms of mixing of states in the tetragonally distorted CuO$_4$Cl$_2$ octahedra together with vibronic coupling. The effect of phonons is revealed as a narrowing and blue-shift of the absorption bands upon cooling~\cite{Lever1968,Burns1993,Solomon1999}; the temperature dependence of the local $d$-$d$ transitions is discussed in Ref.~\cite{SM}.

In this regard, we associate the lowest energy feature observed in experiment to
transitions from $z^{2}$ to  $x^{2}-y^{2}$ while the slightly higher energy
processes correspond to transitions from $t_{2g}$ states
to $x^{2}-y^{2}$. The broad Gaussian peak in Fig.~\ref{sigma_VIS-UV} (a) can be associated to the fact that the $\{xz,yz\}$ levels are almost degenerate as we show in Fig.~\ref{theory_d-d} (c), in addition to temperature effects. Although calculated only for $\mathrm{ZnCu_{3}(OH)_{6}Cl_{2}}$, 
the similarity of optical spectra suggests likewise results for $\mathrm{Y_{3}Cu_{9}(OH)_{19}Cl_{8}}$.

To conclude, with this study we unveil the nature of optical excitations in the frustrated
kagome-lattice Mott insulators $\mathrm{ZnCu_{3}(OH)_{6}Cl_{2}}$ and
$\mathrm{Y_{3}Cu_{9}(OH)_{19}Cl_{8}}$\ by means of optical spectroscopy and
density functional theory calculations. We identify the edge of the
charge-transfer band at 3.3 and 3.6 eV in the experimental spectra, respectively, in good
agreement with calculations corresponding to transitions from the ligand states
to the Cu-$d$ upper Hubbard band with an on-site Coulomb interaction of
8 eV. Inside the gap, we observe dipole-forbidden $d$-$d$ transitions with
small intensity. These local excitations imply only minor changes of the
on-site charge and remain, therefore, rather unaffected by correlation
effects. We thus identify a general route how to handle electronic correlations in
theoretical calculations of optical excitations. Specifically, on-site
Coulomb repulsion $U$ is essential to describe the charge transfer process 
while it plays almost no role for local excitations. Finally, our findings demonstrate that the electrodynamic response of Herbertsmithite and its analogues is similar in nature to the parent compounds of high-$T_{c}$ cuprates, being charge-transfer rather than Mott insulators~\cite{Zaanen1990,Olalde-Velasco2011,MorettiSala2011,Pisarev2010}.

\acknowledgements The project was supported by the Deutsche
Forschungsgemeinschaft (DFG) through grants SFB/TR 49. I.I.M. was supported by A. von Humboldt foundation and by ONR through the NRL basic research program. A.P. and Y.L. equally contributed to this work.

\end{document}